\documentclass[pre,preprint,doublespaced,aps,apssymb,showkeys,showpacs,twocolumn]{revtex4}
\usepackage{amsmath,amssymb}

\usepackage{graphicx}
\usepackage{dcolumn}
\usepackage{bm}
\usepackage{color}

\newcommand{\ber}{\begin{eqnarray}}
\newcommand{\eer}{\end{eqnarray}}
\newcommand{\bea}{\begin{equation}}
\newcommand{\eea}{\end{equation}}

\begin{document}

\title{Conversion of heat to work: An efficient inchworm}
\author{Mayank Sharma}\thanks{mayank.sharma@students.iiserpune.ac.in}
\author{ A. Bhattacharyay}

\email{a.bhattacharyay@iiserpune.ac.in}

\affiliation{Indian Institute of Science Education and Research, Pune, India}

\begin{abstract}
The paper discusses the natural emergence of directed motion in a dimer system due to a structural symmetry breaking. A generalised solution is obtained for the transport of such a system which is driven entirely by bath fluctuations. The result shows the existence of possibility of ratcheting driven by bath fluctuations. If this component of energy conversion driven by bath is taken into account the high efficiency of molecular motors as opposed to paradigmatic ratcheting models can probably be explained.
\end{abstract}
\maketitle      

Various motor proteins like {\it Kinesin} and {\it myosin} that walk on molecular rails (microtubules) inside living cell are essential for bio-mechanical activities, transport etc. Ratcheting models are proposed for such molecular motors \cite{hua, rei}. But, the efficiency of the ratcheting models generally are much smaller than the actual efficiency of the motor proteins they model \cite{astu}. Obviously, the actual biological systems are much more complex than the simple paradigmatic models, however, a large mismatch in the efficiency in general indicates some missing element in these ratchet models. We try to identify one of these possible missing elements to be the work done by the thermal fluctuations of the bath in which ratcheting can happen without any forcing.

An inchworm model for ratcheting is considered in the literature under varied circumstances \cite{stra, ciu, maha, aby, bran, shi, sang, flec}. It has already been shown that an inversion symmetry broken inchworm when immersed in a heat bath can show directed motion because of differential response at the two ends under uniform driving by the equilibrium fluctuations of the bath \cite{ari1}. However, in equilibrium, this kind of a transport although can exist, but, cannot be utilized for any energy conversion from thermal to mechanical form. Any process of energy conversion from such a transport will drive the system out of equilibrium by causing a fall in the local temperature and that will set in other processes of heat and material transport in the bath resulting in entropy production. Obviously, in an isolated heat bath, energy conversion (from thermal to mechanical) by such a transport can only remain through a transient non-equilibrium period of time so long the system reaches a new equilibrium where the transport ceases to exist. 

A non-equilibrium steady state can be imagined for such a system and bath combination which is not isolated from its environment. The bath is thermally connected to a bigger bath (surroundings) which replenishes it all the time by trying to bring it in equilibrium with it. For example, imagine a water tank kept in sun. Here, one can consider a water tank is the bath and the tank is exposed to the sun which establishes its coupling to the bigger environment. The exposure to sun can keep the water tank at the same temperature (over a larger time scale) despite the bath (water tank) losing energy to a system immersed in it. An inchworm can convert energy from the fluctuations of the bath in such a situation in a steady manner. In the present paper we elaborate on such a model of ratcheting where an on-off driving is not present, but, the fluctuations of the heat bath maintained at a constant temperature can drive an inchworm up a potential ramp to convert thermal energy to mechanical energy. The process is of course a non-equilibrium one.
\par
Two paradigmatic ratchet models, namely flashing and rocking ratchet \cite{astu, hadd, han, ethi, hwa, wood} models always involve an intermittent forcing to drive the system (modeled as a single particle) out of equilibrium. The work done by the system in this case is an amount of work extracted mostly from the source of direction-less forcing. There always is present an inversion symmetry breaking of space in such systems at mesoscopic scales which results in the directed transport. The thermal noise assists in this directed motion in the presence of the inversion symmetry breaking by providing the necessary spread in the position distribution at the scales comparable to the scale of the inversion symmetry breaking. However, in these models, the thermal energy of the bath does not appreciably contribute to the energy gain (efficiency) of the system, rather, restricts the efficiency of such systems. However, if thermal fluctuations of the bath can drive a system in a direction, it will directly contribute to the efficiency of ratcheting system. The purpose of the present work is to underline such a possibility.
\par
The requirement of the intermittent forcing of the above mentioned two paradigmatic ratchet models are normally played with to increase efficiency \cite{gao}. These driving processes are subtle and depend on many other parameters of the model system to enhance efficiency \cite{astu}. Obviously, such a supply of energy is needed to maintain a steady state when energy is being extracted from the system. The present model shows that, maintaining the steady state would obviously need external energy supply to the bath, but, that without any subtle adjustment of intermittent forcing. Imagine that, if such a ratcheting is happening inside a living cell, the temperature of the cell is primarily contributing to the efficiency of the ratcheting where the intermittent active processes for any other purpose are helping maintain the temperature of the cell. Note that, when we mention of a temperature in the present case, we mean that the non-equilibrium processes are happening at a smaller time scale and the temperature is measured over a relatively larger time scale.
\par
In what follows, we consider an inchworm model consisting of two particles of varied damping coefficients i.e., the particles are structurally different and, hence, their responses to the bath fluctuations are also different. The damping coefficients are considered to be functions of the distance between the particles. The reason for this is that, the number of the bath degrees of freedom the particles are exposed to are functions of the relative distance between the particles. Moreover, the particles can talk to each other by any attractive force and cannot pass through each other because of a sufficient potential barrier which the thermal energy can not overcome. Such a system when confined to one-dimension has a broken inversion symmetry which remains intact in the absence of rotations which is not allowed in one-dimension. 
\par
Since this modeling involves multiplicative noise, we follow here the standard It\^o convention which results in a modified Boltzmann distribution of the internal coordinates \cite{ari2,ari3}. In the following, we first present the general model with arbitrary potentials and damping and analyze it to show that in general there always exists directed motion of such a symmetry broken system up against a constant force. Then we take a special case of simply solvable potentials to get the relevant numbers which we compare with the numbers obtained from a direct numerical simulation of the system and present results. We give a general discussion on the usefulness of such a process to end the paper.
\section{general model}
The over-damped equations for particles of unit mass  at positions $x_1$ and $x_2$ where $x_1 > x_2$ are

\begin{align}
\begin{split}
 & \frac{d x_{1}}{d t}  = \frac{(F-\frac{\partial V(z)}{\partial x_{1}})}{\gamma_{1}\Theta(z-z_{min})+\mu_{1}[1-\Theta(z-z_{min})]} 
  \\ \\  +&\eta_{1}(t) \sqrt \frac{2k_{B}T}{\gamma_{1}\Theta(z-z_{min})+\mu_{1}[1-\Theta(z-z_{min})]} 
  \end{split}
 \end{align}
 
 and

\begin{align}
\begin{split}
 & \frac{d x_{2}}{d t}  = \frac{(F-\frac{\partial V(z)}{\partial x_{2}})}{\gamma_{2}\Theta(z-z_{min})+\mu_{2}[1-\Theta(z-z_{min})]} 
  \\ \\  +&\eta_{2}(t) \sqrt \frac{2k_{B}T}{\gamma_{2}\Theta(z-z_{min})+\mu_{2}[1-\Theta(z-z_{min})]}.
  \end{split}
\end{align}

Here, $F$ is a constant force against which the system of particles will work. The internal coordinate of the system is $z=x_1-x_2$. Particles are bound by an attractive potential and, at short distances, there acts a repulsive potential which prevents the particles from passing through each other. Both the attractive and the repulsive parts of the potentials are encoded by the function $V(z)$. The symmetry of the system is broken by the different damping constants $\mu_i$ and ($i=1,2$) which works below a distance $z<z_{min}$ and $\gamma_i$ working at and above the cut off distance $z\geq z_{min}$. The $\Theta(z-z_{min})$ is a theta function which is unity for the argument greater than and equal to zero and is zero otherwise. The cutoff distance $z_{min}$ is set such that the damping during the collision between particles becomes different than that when the particles are far apart from each other. This is reasonable because the number of bath degrees of freedom the particles will be exposed to are different when they are in contact with each other than then they are a distance apart. The temperature of the bath (as measured over a larger time scale) is given by $T$ and $k_B$ is Boltzmann constant. The unit strength Gaussian white noises are $\eta_i$.
\par
We are going to analyze this system to see if it can move against the global force $F$ under the broken symmetry conditions in a steady state of the bath defined by its temperature $T$. The diffusivity considered here to be $D_i= \frac{k_{B}T}{\gamma_{i}\Theta(z-z_{min})+\mu_{i}[1-\Theta(z-z_{min})]}$ which is in accordance with the Stokes-Einstein relation. Obviously the time scale of the dynamics that we are considering here is larger than the time scale over which the temperature and the diffusivity gets defined and, therefore, it is an over-damped dynamics.
\par
Let us go to the centre of mass $x=\frac{x_1+x_2}{2}$ and internal coordinate $z$ which yields the dynamics

\begin{align}
\begin{split}
 \frac{d x}{d t} & =\frac{F}{2}\left [ \frac{1}{\Gamma_1(z)}+\frac{1}{\Gamma_2(z)} \right]+\sqrt{k_{B}T/2} \xi_{1}(z,t)
 \\ &-\frac{1}{2}\frac{\partial V(z)}{\partial z}\left [ \frac{1}{\Gamma_1(z)}-\frac{1}{\Gamma_2(z)} \right]
 \end{split}
\end{align}

and

\begin{align}
\begin{split}
 \frac{d z}{d t} &= \left [F\frac{\Gamma_2(z)-\Gamma_1(z)}{\Gamma_2(z)+\Gamma_1(z)}-\frac{\partial V(z)}{\partial z}\right ] \\ & \times\left [ \frac{1}{\Gamma_1(z)}+\frac{1}{\Gamma_2(z)} \right ] +\sqrt{2k_{B}T}\xi_{2}(z,t)
\end{split}
\end{align}

where \newline
$\Gamma_i(z) = {\gamma_{i}\Theta(z-z_{min})+\mu_{i}[1-\Theta(z-z_{min})]}$. Stochastic forces are $\xi_1=\frac{\eta_1(t)}{\sqrt {\Gamma_1(z)}}+\frac{\eta_2(t)}{\sqrt {\Gamma_2(z)}}$ and $\xi_2=\frac{\eta_1(t)}{\sqrt {\Gamma_1(z)}}-\frac{\eta_2(t)}{\sqrt{\Gamma_2(z)}}$ which have zero mean.
\par
Important to note here that the internal coordinate ($z$) has a dynamics which is independent of the center of mass and can be solved to get the distribution of the internal coordinate. The Smoluchowski equation resulting from the standard Kramers-Moyal expansion for the internal coordinate (Eq.(4)) is of the form
\begin{align}
\begin{split}
\frac{\partial \rho(z,t)}{\partial t}& = - \frac{\partial}{\partial z}\left [h(z)\rho(z,t) - \frac{\partial D(z)\rho(z,t)}{\partial z} \right ]\\
&= -\frac{\partial J(z,t)}{\partial z}.
\end{split}
\end{align}

In the above equation  $\rho(z,t)$ is the probability density,
$h(z) = \left (F\frac{\Gamma_2(z)-\Gamma_1(z)}{\Gamma_2(z)+\Gamma_1(z)}-\frac{\partial V(z)}{\partial z}\right )\left (\frac{1}{\Gamma_1(z)}+\frac{1}{\Gamma_2(z)}\right )$ is the drift velocity and $D(z)=k_BT\left ( \frac{1}{\Gamma_1(z)}+\frac{1}{\Gamma_2(z)}\right )$ is the effective diffusivity for the internal coordinate.

\par 
 Considering an equilibrium solution $\rho(z)$ which means $J(z,t)=0$ and is allowed for the internal coordinate distribution even in the presence of a global force because of the decoupling of the centre of mass and internal coordinates we get
\begin{align}
    \begin{split}
& \rho(z)=\\ & \frac{N}{D(z)}\exp{\frac{\int_{0}^z {dz^\prime\left (F\frac{\Gamma_2(z^\prime)-\Gamma_1(z^\prime)}{\Gamma_2(z^\prime)+\Gamma_1(z^\prime)}-\frac{\partial V(z^\prime)}{\partial z^\prime}\right )}}{k_BT}},
\end{split}
\end{align}
where $N$ is a normalization factor.

The stationarity of the distribution is ensured also by the fact that 
\begin{align}
\begin{split}
& \langle \frac{dz}{dt}\rangle =  N \times \\ & \int_{0}^{\infty}{dz\frac{d}{dz}\exp{\frac{\int_{0}^z {dz^\prime\left (F\frac{\Gamma_2(z^\prime)-\Gamma_1(z^\prime)}{\Gamma_2(z^\prime)+\Gamma_1(z^\prime)}-\frac{\partial V(z^\prime)}{\partial z^\prime}\right )}}{k_BT}}}\\ &\equiv 0
\end{split}
\end{align}
under the consideration that the Boltzmann factor $\exp{\left ({\int_{0}^z {dz^\prime\left (F\frac{\Gamma_2(z^\prime)-\Gamma_1(z^\prime)}{\Gamma_2(z^\prime)+\Gamma_1(z^\prime)}-\frac{\partial V(z^\prime)}{\partial z^\prime}\right )}}/{k_BT}\right )}$ vanishes at $z=0$ and at $z=\infty$. Let us emphasize on the fact that, had we not taken the probability density in the form shown above and used a Boltzmann distribution instead by dropping the $z$ dependent amplitude for a constant normalization factor we would have not got this vanishing average $z$-current.
\par
Now, we are interested in finding out the average center of mass velocity as given by

\begin{align}
\begin{split}
 v=& \langle \frac{d x}{d t}\rangle   = \int_0^{\infty}{dz\rho(z)\frac{F}{2}\left [ \frac{1}{\Gamma_1(z)}+\frac{1}{\Gamma_2(z)} \right ]}
  \\ &-\int_0^\infty{dz\rho(z)\frac{1}{2}\frac{\partial V(z)}{\partial z}\left [ \frac{1}{\Gamma_1(z)}-\frac{1}{\Gamma_2(z)} \right ]}.
  \end{split}
\end{align}

Notice that, the first term on the r.h.s., of the above equation is always going to produce a center of mass motion downhill the global potential and the second term on the same side would only produce a center of mass motion when there is the structural symmetry breaking present i.e. $\Gamma_1(z) \neq \Gamma_2(z)$. By the adjustment of the values of $\Gamma_1(z)$ and $\Gamma_2(z)$ (or, more physically,  the Stokes radii  of the particles) we can make this part of the center of mass motion to counteract or even overcome that due to the global force and make the system move  against the global force. In such a situation, the thermal energy of the bath will get directly converted to the potential energy of the system and that is precisely the matter of our interest here. We will look into this model by considering specific forms for $V(z)$ and $\Gamma_i(z)$ in the following. The efficiency of the ratchet in working against the force $F$ will be defined as $\eta = \frac{| v F|}{k_BT}$. \\  Our results would show that there can exist within the over-damped (slow dynamics) limit a wide range of parameters to make the system move against the global force. 

\section{Numerical results}
In this section, we are going to present various results obtained by simulation of the model given by eqn.(1) and eqn.(2) and compare those with the theoretical ones. The inter particle interaction potential that we have considered is $V(z) = \frac{\alpha}{2}(z-z_{min})^2$, where $\alpha$  is the spring constant. We have implemented Euler-Maruyama scheme to simulate a pair of overdamped stochastic differential equations.
The discrete time step $\Delta t=10^{-4}$, total number of time steps = $10^{7}$ are chosen for all the simulations.  Fig.1 shows the system in absence of any external force. Fig.2 shows the same system in a the presence of an external force F$<$ $F_{stall}$ where $F_{stall}$ is the force that stalls the motion against it. Fig.3 , Fig.4 and Fig.5 show the variation of average velocity of centre of mass and efficiency with respect to various parameters of the system.

The effective potential $V_{eff}$ experienced by either particle is defined as 
$V_{eff}=\frac{\int{dz[\frac{dV}{dz}+(-1)^iF}]}{\Gamma_{i}}$with $i$ being the particle index.
The probability density is discontinuous about $z_{min}$  because of introduction of  theta function in the model, breaking symmetry of damping coefficients about $z_{min}$. The  center of mass  drifts freely in the absence of an external force because of symmetry broken damping coefficients for $z<z_{min}$  and $z$ $\geq $ $z_{min}$. One obtains a constant average centre of mass velocity in the model under symmetry broken conditions. The cause of a  constant average centre of mass velocity  is  the constant balance between the effective symmetry broken harmonic force and overall damping on the system. 

We set the constant parameters to be $\alpha = 100$, $ k_B T =1$, $z_{min}=1$, $\gamma_{1}=1$,  $\gamma_{2}=2$, $\mu_{1}=4$, $\mu_{2}=3$ if not otherwise mentioned and for Fig.(1) $F=0$. Where the figures 1(a) and 1(b) are showing the distribution of the internal coordinate and the effective potentials as seen by the particles, the position of the center of mass $x$ is plotted against time in sub-figure (c). Fig. 1(d) shows the center of mass trajectory in the case where the symmetry has not been broken i.e., the $\gamma$s and the $\mu$s are the same. One can see that the directed motion of the center of mass is lost in the presence of no broken symmetry. On average over a very large time and different initial conditions, it has been checked that the centre of mass does not move in any particular direction on average.

\begin{figure}

\includegraphics[scale=0.14]{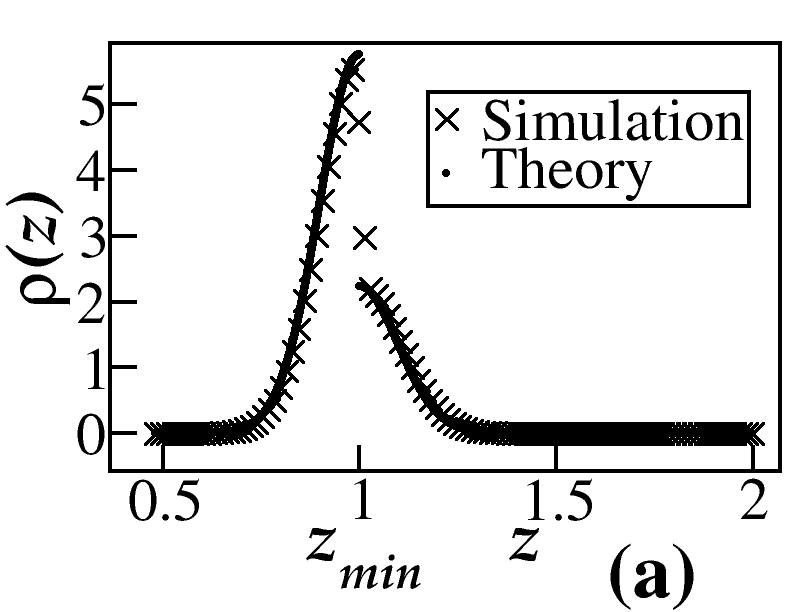}
\includegraphics[scale=0.14]{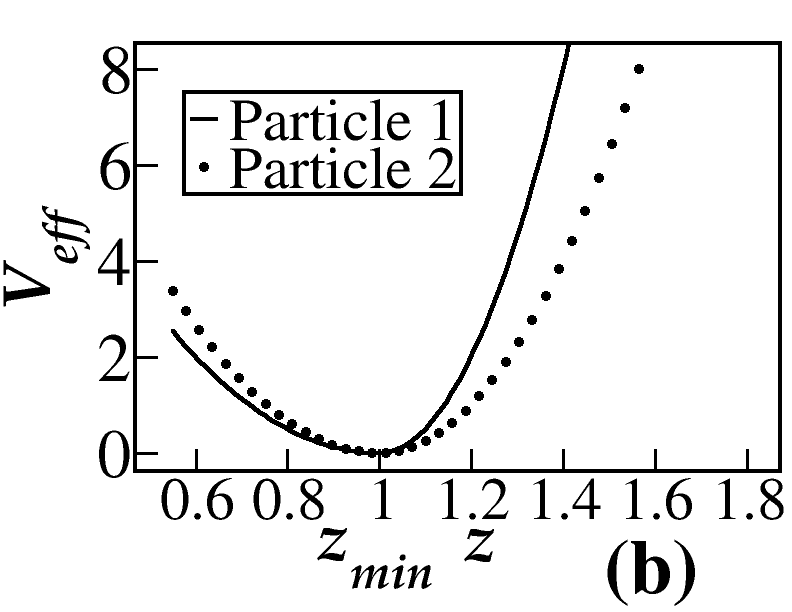}
\includegraphics[scale=0.14]{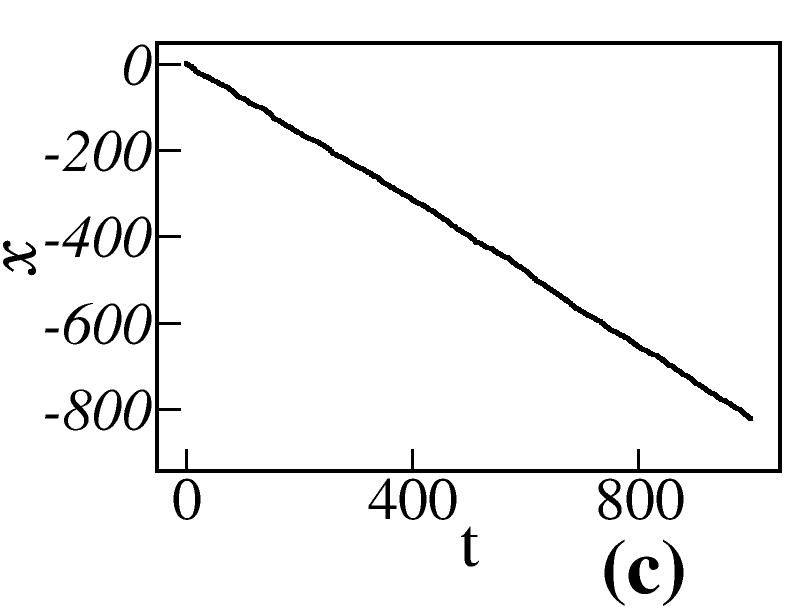}
\includegraphics[scale=0.14]{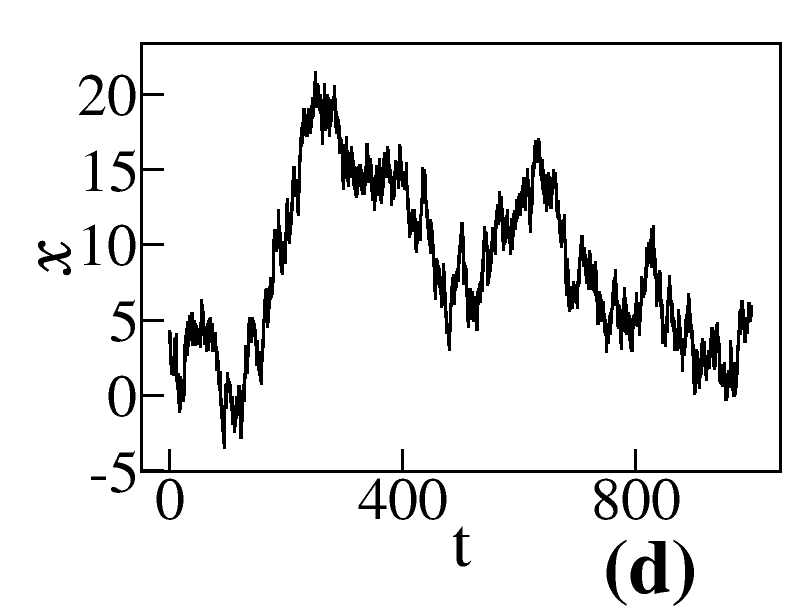}
\caption{ { \footnotesize Probability density, effective potential and  motion of centre of mass being shown for usual parameters in (a), (b) and (c) respectively in absence of any external force.  Sub-figure (d) : depicting the motion of centre of mass (non-directed) for symmetrical damping coefficients: $\gamma_{1}=1,\gamma_{2}=1 ,\mu_{1}=3,\mu_{2}=3.$} 
}       
\end{figure}

In Fig.2, we keep all the parameters the same as Fig.(1) except for $F=1.5$. Fig.2 shows similar graphs as Fig.1 but, now in the presence of a force applied in a direction opposite to the direction of motion of the system. The trajectories in 2(c) is against the force. Fig. 2(d) shows the situation where the symmetry is not broken just as in a parameter setting as in Fig.1(d), however, in the presence of a force the center of mass is simply moving in the direction of the external force and as a result the direction of motion is opposite to that of symmetry broken ratcheting dimer.

\begin{figure}

\includegraphics[scale=0.14]{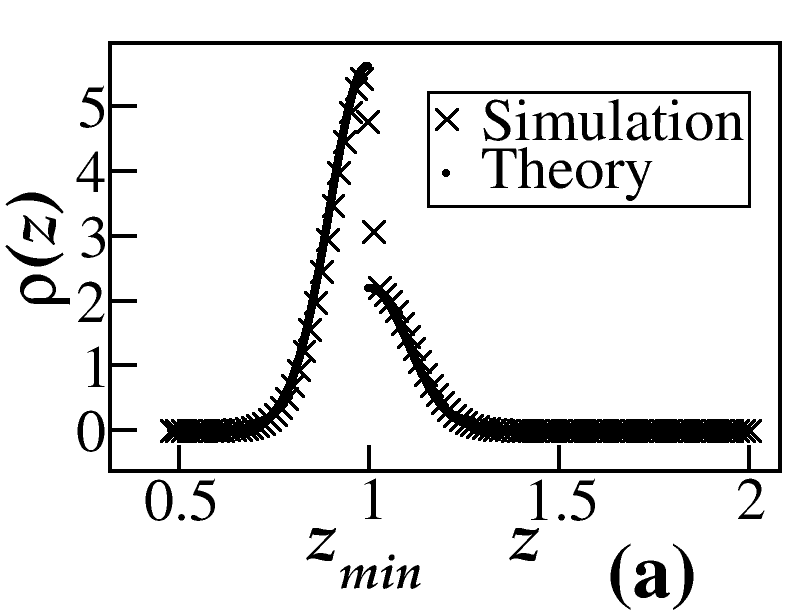}
\includegraphics[scale=0.14]{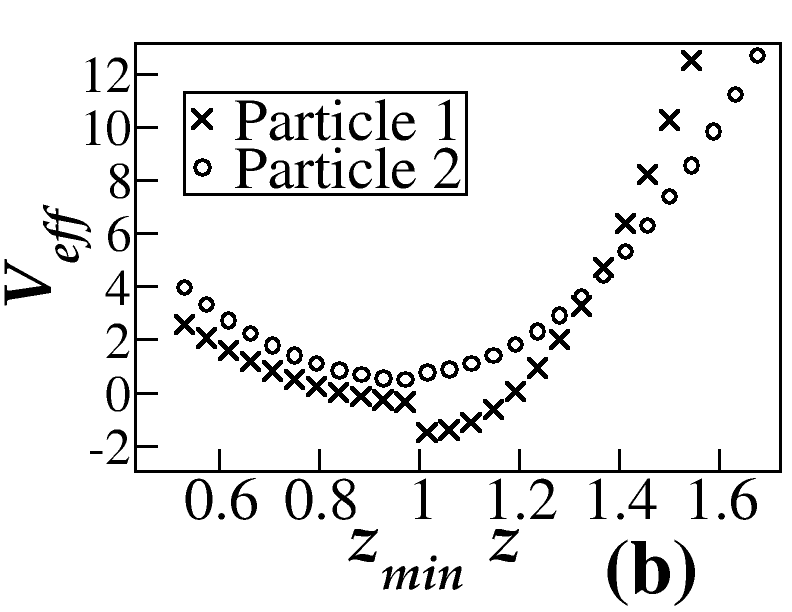}
\includegraphics[scale=0.14]{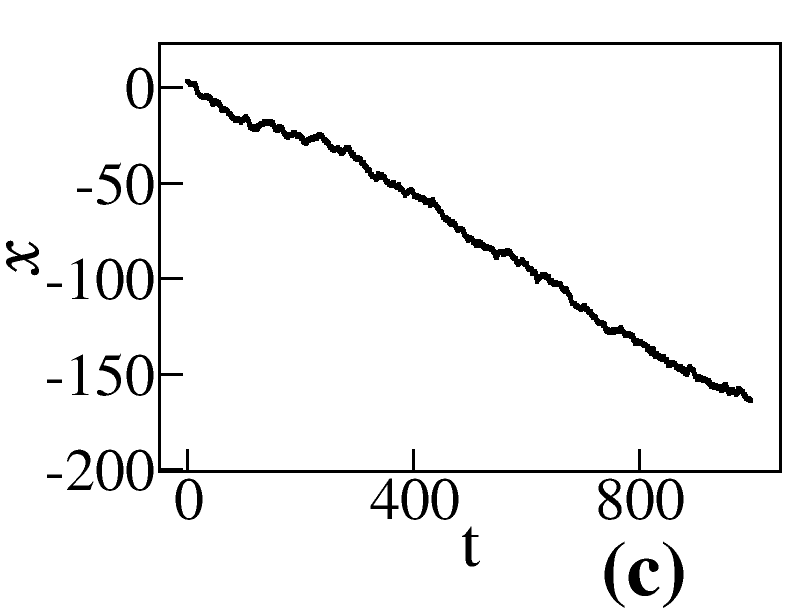}
\includegraphics[scale=0.14]{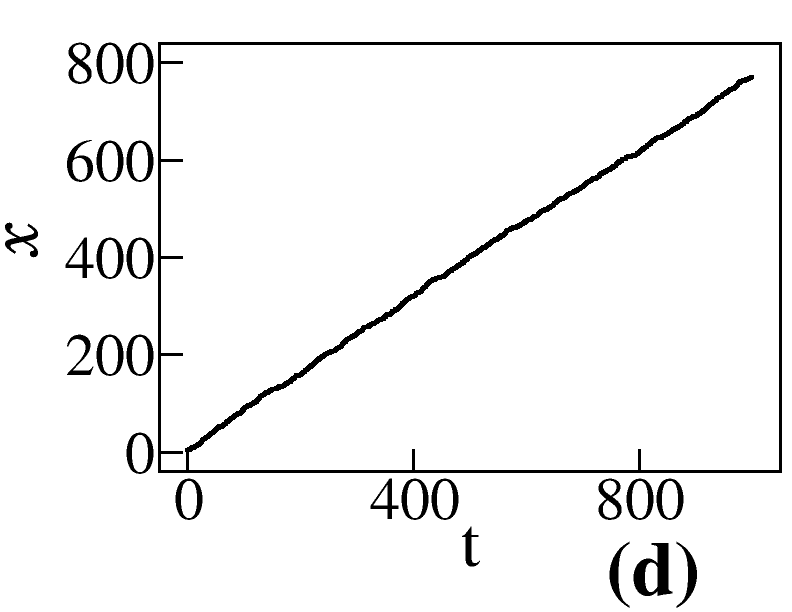}

\caption{ \footnotesize Probability density , effective potential and  motion of centre of mass being shown for usual parameters in (a) , (b) and (c) respectively  for a non zero external force . Sub-figure (d) : depicting the motion of centre of mass  for symmetrical damping coefficients : $\gamma_{1}=1,\gamma_{2}=1 ,\mu_{1}=3 ,\mu_{2}=3.$ 
}  

\end{figure}

\begin{figure}

\includegraphics[scale=0.14]{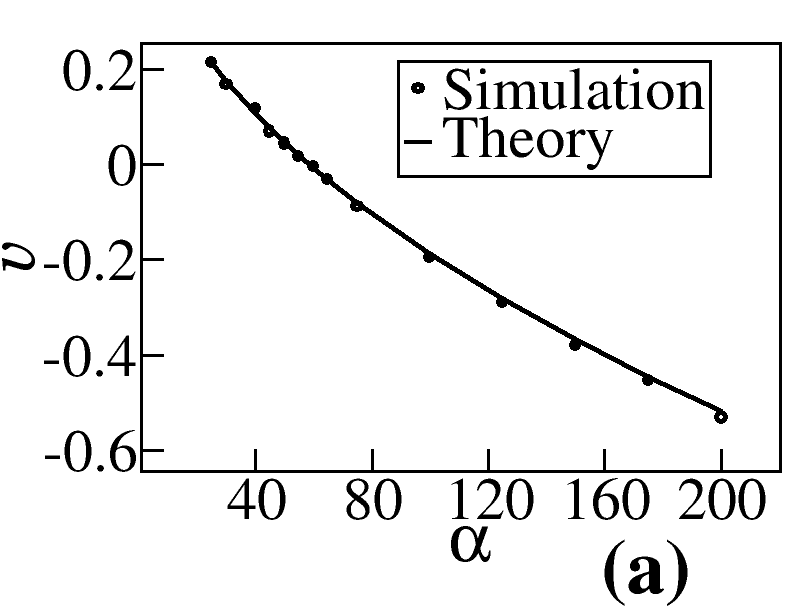}
\includegraphics[scale=0.14]{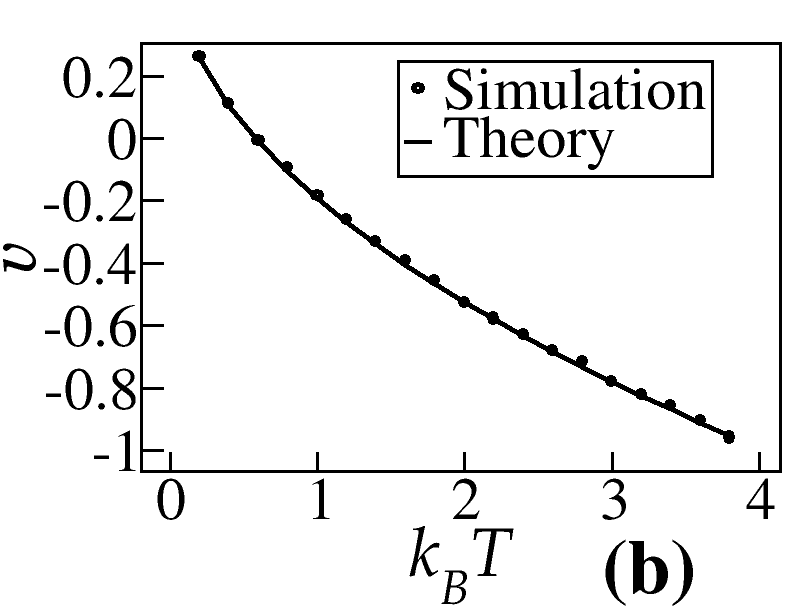}
\includegraphics[scale=0.14]{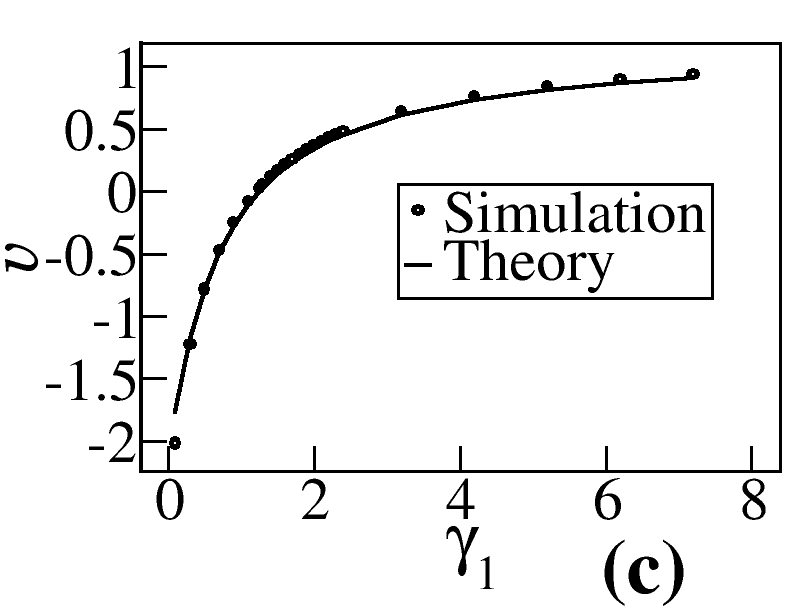}
\includegraphics[scale=0.14]{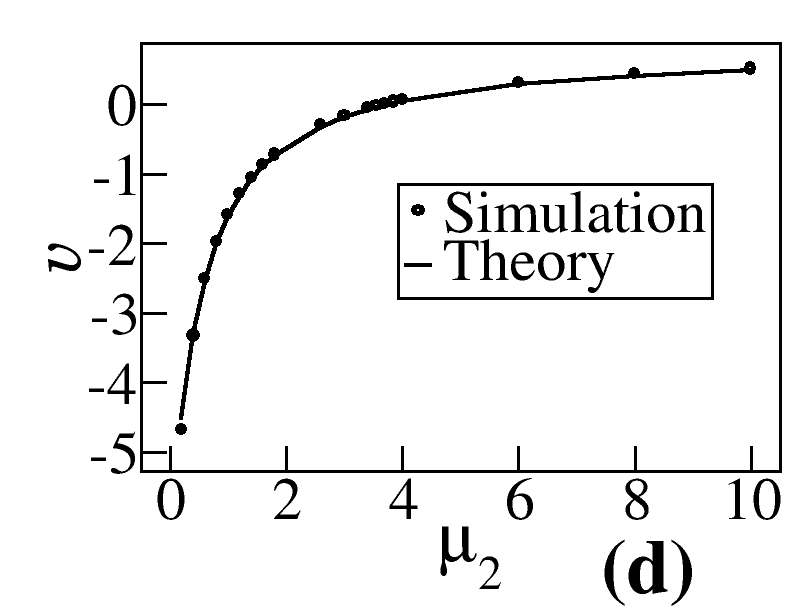}

\caption{ \footnotesize  Average  centre of mass velocity plotted against various parameters of the system. In (b),  $z_{min}=100$ and in rest of sub-figures, $z_{min}=1$ as usual.
}       
\end{figure}
Fig.3 shows the variation of average velocity of centre of mass  with respect to various parameters of the system. In Fig.3, we set the constant parameters to be $\alpha = 100$, $ k_B T =1$, $z_{min}=1$, $\gamma_{1}=1$,  $\gamma_{2}=2$, $\mu_{1}=4$, $\mu_{2}=3$ and $F=1.5$ as before. The  $\alpha$ is coupling parameter, as  $\alpha$ increases the  potential well becomes steeper. The natural tendency of F is to make system move in positive x direction and that of  symmetry broken internal force is to move it in negative x direction. As $\alpha$ increases, the strength of symmetry broken internal force becomes more. Eventually the system's average velocity in the direction of the external force starts decreasing which becomes zero at $\alpha=58.6$ for the given strength of the external force and then with further increase in  $\alpha$, the velocity of the system against the force in negative x direction keeps on increasing. However, one should keep in mind that one is not allowed to increase the $\alpha$ to an arbitrarily high value because that would breach the condition of slow over-damped dynamics followed in the present analysis. We use the value of the efficiency $\eta$ as a check for the system remaining in over-damped regime. So, long $\eta$ is well below unity, we consider the parameter values are good for over-damped description and $\eta \simeq 1$ indicates a breakdown of the slow dynamics. 

Fig.3(b) shows the average velocity of the center of mass against the thermal energy $k_BT$. The choice of $z_{min}=100$ for this graph is to avoid crossing over of particles at high temperatures. Changing $z_{min}$ just translates the  potential along z axis. Thus, efficiency  $\eta$ and average centre of mass velocity $v$ are independent of choice of $z_{min}$. The average velocity becomes zero at $k_{B}T = 0.586$ due to the contribution of $F$ is just balanced there by ratcheting. As temperature increases, the ratcheting overcomes external forcing and drives the system against the external force. In Fig.3(c), the average centre of mass velocity  becomes zero at $\gamma_{1}\simeq 1.25$. In Fig.3(d), we see that average velocity of centre of mass overcomes the external force at $\mu_{2}\simeq 3.75$. The divergence of the average velocity towards the smaller values of $\gamma_i$ and $\mu_i$ is a result of the increase in the difference, in $|\gamma_1-\gamma_{2}|$ and $|\mu_1-\mu_2|$. The increase in these differences means the shortening of the time scales and a possible breakdown of the slow dynamics if it drives the system beyond the efficiency limit (actually Carnot limit) set by us. 

\begin{figure}

\includegraphics[scale=0.14]{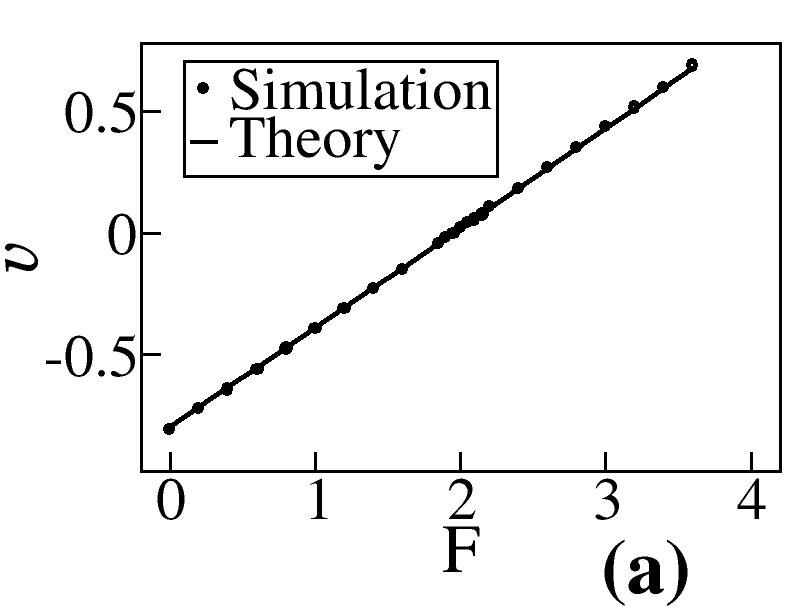}
\includegraphics[scale=0.14]{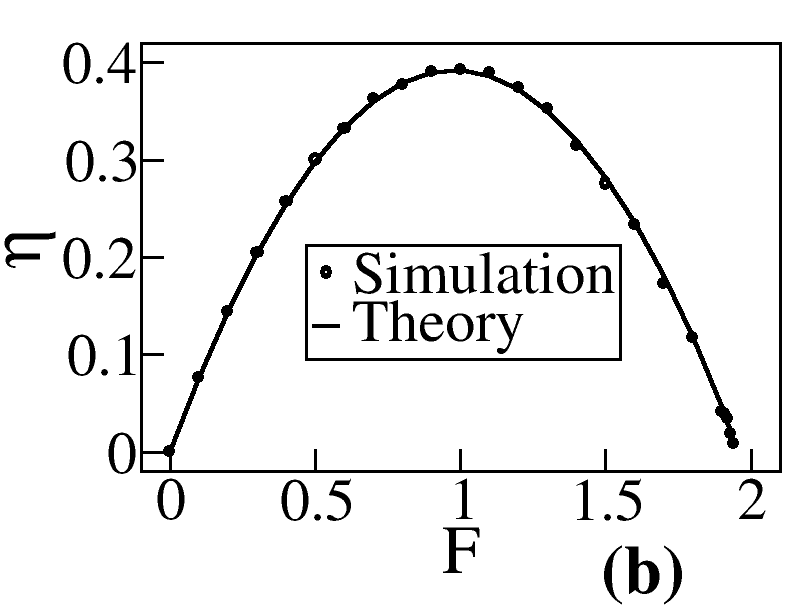}

\caption{ \footnotesize  Results of variation of average velocity of center of mass and efficiency  shown against external force. }
\end{figure}

In Fig.4(a) the external force $F$ is varied keeping all other parameters fixed  to the values already mentioned to look at the effects of the external force on the average center of mass velocity and efficiency of the system. As $F$ increases, it opposes the motion in negative x direction and the average velocity of the system decreases till it reaches zero velocity and eventually the system moves in positive x-direction. The average velocity of centre of mass of the system becomes zero at about $F=1.96$, so the stall force $F_{stall}$ is about 1.96. The Fig.4(b) shows a plot of the efficiency $\eta$ against the variation in external force. We have chosen the maximum external force up to the $F_{stall}$. The maximum efficiency against the external force is about $40 \%$ obtained at about half of the $F_{stall}$.

\begin{figure}

\includegraphics[scale=0.14]{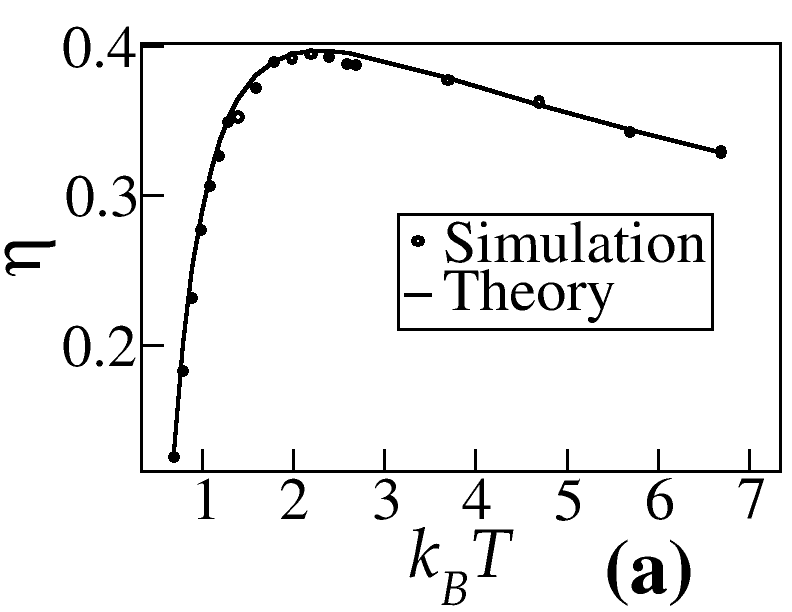} 
\includegraphics[scale=0.14]{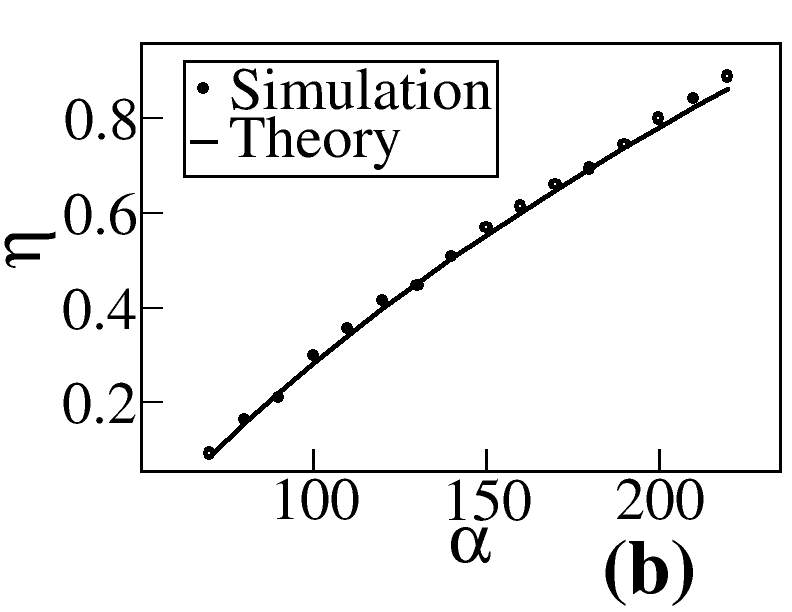}
\includegraphics[scale=0.14]{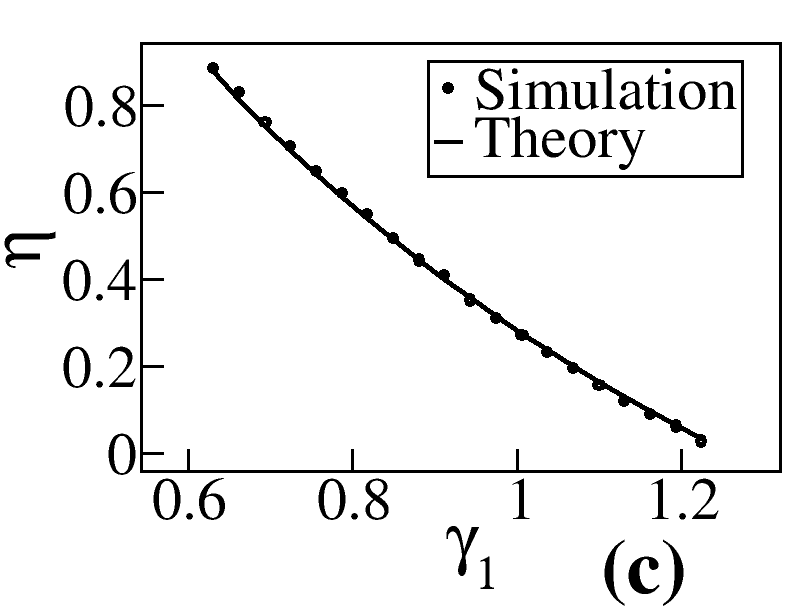}
\includegraphics[scale=0.14]{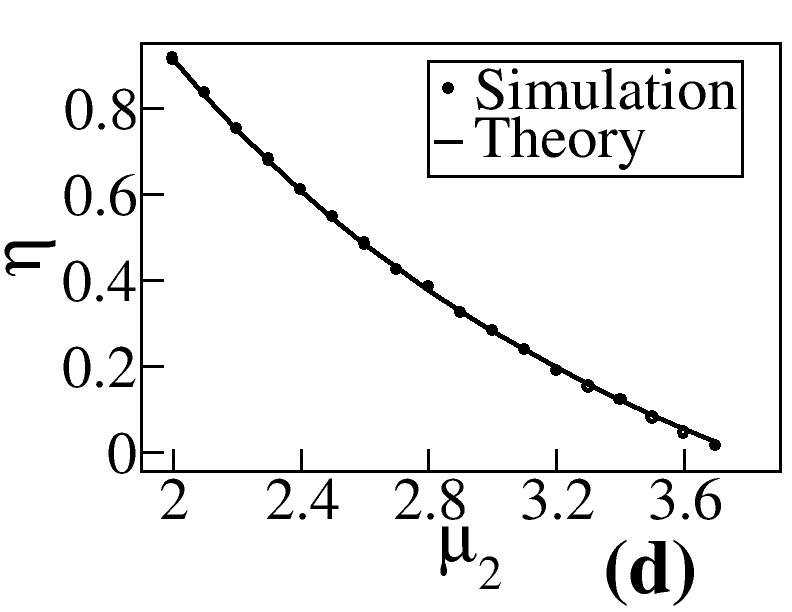}

\caption{ \footnotesize The variation of efficiency of system with various parameters of the system. In (a), $z_{min}=100$  and in rest of the sub-plots $z_{min}=1$ as usual.
}
\end{figure}

In Fig.5 we set the same constant parameters as for Fig.3 unless 
 mentioned otherwise. In Fig.5(a), we see the variation of efficiency with respect to temperature. We  start our counting for efficiency from those values of $k_B T$ where the system moves against global force. The efficiency attains a maximum value of about 0.396 at $k_{B}T= 2.4$.

 In Fig.5(b) we show the variation of efficiency with the spring parameter. The efficiency increases with spring parameter and eventually would cross the unity at large $\alpha$. These higher values of $\alpha$ are obviously indicative of a breakdown of the over-damped theory and we restrict ourselves to the values much smaller than those. Note that, $\alpha$ represents an inverse square of characteristic time scale for unit mass. Therefore, very large $\alpha$ means a very small time scale. However, the over-damped dynamics being necessarily slow, one cannot go to a very large $\alpha$ without violating standard theory.

 In Fig.5(c) and Fig.5(d) we show the variation of efficiency with other two parameters $\gamma$ and $\mu$. The difference in $\gamma_i$s and $\mu_i$s are the measure of broken symmetry. As a result, the efficiency increases/decreases with the increase/decrease of these differences.

\section{discussion}
In the present paper we have investigated an inchworm structured ratchet which due to its structural asymmetry is able to filter bath fluctuations to result in directed motion against a force. When energy is extracted from such a process, it is necessarily a non-equilibrium one. In our analysis, we have employed an over-damped model which is necessarily a slow one. We have taken this path, which is quite common in modeling such systems, because for a slow process even in non-equilibrium conditions one can talk of thermodynamic quantity like temperature.

The present analysis clearly shows that there can exist a purely thermal component of energy originating from the fluctuations of bath which contributes to the power of such a ratchet. In view of the fact that standard flashing and rocking ratchet have their efficiency almost totally dependent on the external forcing and not the bath fluctuations \cite{astu}, identification of the present mechanism is particularly interesting. The present mechanism, when combined with a flashing or rocking ratchet, can account for an enhanced efficiency in a very general manner. The whole extra efficiency would result in this case from the consideration of the structure of the moving object which conventionally is considered to be a point particle in flashing and rocking ratchets. This is a very reasonable consideration that the structure matters which gets effectively ignored when the inversion symmetry is broken by the presence of a ratchet (saw-tooth like) potential over space on which a single particle moves.

The present analysis also in principle shows the possibility of conversion of, for example, solar energy that heats up a bath into conserved work by the use of inchworm like machines. Beyond the scope of partially understanding the motion of motor proteins in biological systems (which involve other complexities), the present model indicates this particular possibility of conversion of heat to work by use of machines which do not require a power stroke. Bath fluctuations under non mechanically driven non-equilibrium conditions are enough to generate storage of useful work.

\end{document}